\documentclass[%
reprint,
superscriptaddress,
amsmath,
amssymb,
aps,
prx,
longbibliography,
floatfix,
]{revtex4-2}

\usepackage{graphicx}
\usepackage{dcolumn}
\usepackage{bm}
\usepackage{hyperref}
\usepackage{xr}
\usepackage{lipsum}
\usepackage{braket}
\usepackage{xcolor}
\usepackage{pict2e}
\usepackage{overpic}
\usepackage{rotating}
\usepackage{longtable}
\newcommand{\cor}[1]{\textcolor{black}{#1}}
\begin{document}
\preprint{APS/123-QED}
\bibliographystyle{unsrtnat} 
\title{Performance of high impedance resonators in dirty dielectric environments}
\author{J.\,H.~Ungerer}
\affiliation{
Department of Physics, University of Basel, Klingelbergstrasse 82 CH-4056, Switzerland
}
\affiliation{
Swiss Nanoscience Institute, University of Basel, Klingelbergstrasse 82 CH-4056, Switzerland
}
\author{D.~Sarmah}
\affiliation{
Department of Physics, University of Basel, Klingelbergstrasse 82 CH-4056, Switzerland
}
\author{A.~Kononov}
\affiliation{
Department of Physics, University of Basel, Klingelbergstrasse 82 CH-4056, Switzerland
}
\author{J.~Ridderbos}
\thanks{Current address: NanoElectronics Group, MESA Institute for Nanotechnology, University of Twente, P.O. Box 217, 7500 AE Enschede, The Netherlands}
\affiliation{
Department of Physics, University of Basel, Klingelbergstrasse 82 CH-4056, Switzerland
}
\author{R.~Haller}
\affiliation{
Department of Physics, University of Basel, Klingelbergstrasse 82 CH-4056, Switzerland
}
\author{L.~Y.~Cheung}
\affiliation{
Department of Physics, University of Basel, Klingelbergstrasse 82 CH-4056, Switzerland
}
\author{C.~Sch{\"o}nenberger}
\homepage{www.nanoelectronics.unibas.ch}
\affiliation{
Department of Physics, University of Basel, Klingelbergstrasse 82 CH-4056, Switzerland
}
\affiliation{
Swiss Nanoscience Institute, University of Basel, Klingelbergstrasse 82 CH-4056, Switzerland
}
\date{\today}

\begin{abstract}
High-impedance resonators are a promising contender for realizing long-distance entangling gates between spin qubits.
Often, the fabrication of spin qubits relies on the use of gate dielectrics which are detrimental to the quality of the resonator.
Here, we investigate loss mechanisms of high-impedance NbTiN resonators in the vicinity of thermally grown SiO\textsubscript{2} and Al\textsubscript{2}O\textsubscript{3} fabricated by atomic layer deposition. 
We benchmark the resonator performance in elevated magnetic fields and at elevated temperatures and find that the internal quality factors are limited by the coupling between the resonator and two-level systems of the employed oxides.
\cor{Nonetheless, the internal quality factors of high-impedance resonators exceed $10^3$ in all investigated oxide configurations which implies that the dielectric configuration would not limit the performance of resonators integrated in a spin-qubit device.}
Because these oxides are commonly used for spin qubit device fabrication, our results allow for straightforward integration of high-impedance resonators into spin-based quantum processors.
Hence, these experiments pave the way for large-scale, spin-based quantum computers.
\end{abstract}
\maketitle
\section{Introduction}
\begin{figure*}[!htb]
    \centering
    \includegraphics[width=\textwidth]{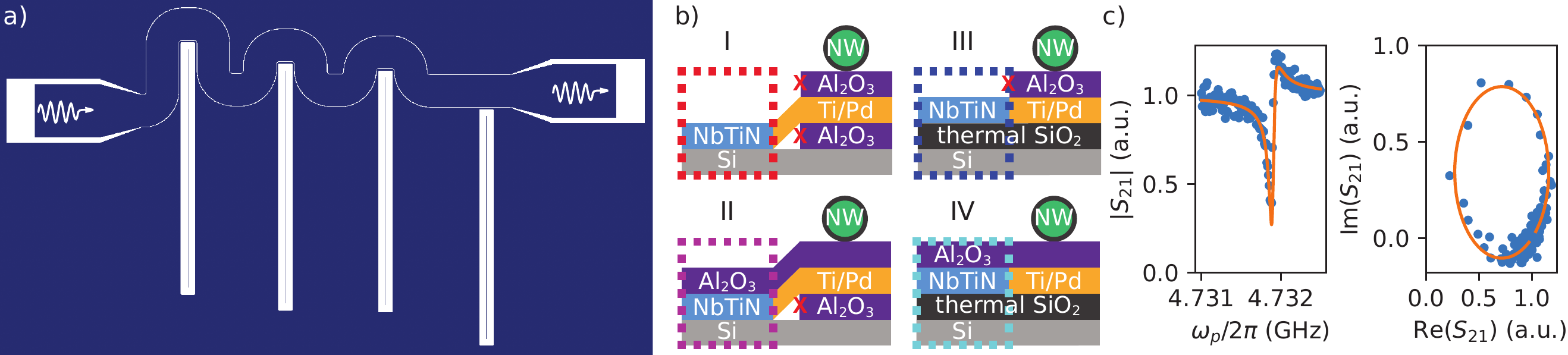}
    \caption{\textbf{Device overview.} a) Top-view design of the four high-impedance ($\sim$ 2 k$\Omega$) half-wave resonators, capacitively coupled to a much wider 50~$\Omega$ matched feedline.
    b) Side-view schematics of four different dielectric configurations that were investigated. The color of the dashed boxes corresponds to their respective configurations throughout the paper.
    The right part of each schematic illustrates how a nanowire device can be realized, given the dielectric configuration on its left.
    c) Amplitude and IQ-excursion of exemplary resonance. The orange line is a fit to the data.}
    \label{fig:hangers}
\end{figure*}
\cor{Understanding the origin of dielectric losses is crucial when exploiting superconducting resonators for quantum information science. But studies of dielectric losses in various dielectrics have so far been based on low-impedance resonators}~\cite{martinis2005decoherence,o2008microwave,gao2008experimental,cicak2010low,pappas2011two,paik2010reducing,mcrae2020dielectric,muller2022magnetic,mcrae2020materials}.\cor{ 
 However, a large resonator impedance is desirable, in-particular in the context of spin-qubits, as the coupling to a weak electric dipole moment scales with the square root of the impedance}~\cite{blais2004cavity}\cor{.
Previous studies showed that high-impedance resonators in a clean electrostatic environment typically reach quality factors on the order of $10^5$}~\cite{singh2014molybdenum,samkharadze2016high,yu2021magnetic,grunhaupt2018loss} \cor{in the absence of dielectrics.}

\cor{In applications involving semiconductor qubits, the quality factors of high-impedance resonators are typically limited to $\sim10^3$ due to gate leakage}~\cite{mi2017circuit,harvey2020chip}\cor{. Despite these relatively low quality factors, high-impedance resonators have realized important breakthroughs such as coherent coupling between a single photon and a single charge qubit}~\cite{stockklauser2017strong,mi2017strong}, coherent spin-photon coupling~\cite{samkharadze2018strong,mi2018coherent,landig2018coherent,yu2023strong,ungerer2023strong} and distant resonant charge-to-charge~\cite{van2018microwave} and spin-to-spin~\cite{borjans2020resonant} coupling as well as rapid-gate based spin readout~\cite{zheng2019rapid} and the demonstration of ultrastrong charge-photon coupling~\cite{scarlino2021situ}.
\cor{Impressively, the implementation of high-impedance resonators with quality factors of $\sim10^3$ 
has enabled} distant virtual-photon mediated charge-to-charge~\cite{van2018microwave} and spin-to-spin~\cite{harvey2022coherent} coupling. 

\cor{Incorporating high-impedance resonators with dielectrics would allow one to use established fabrication recipes that employ gate dielectrics aiming on electrostatic stability of quantum dot devices.
The fact that the quality factor of a high-impedance resonator, incorporated in a real device, is not limited by dielectric losses therefore raises the following question: To what extent can dielectrics be employed when fabricating resonator devices without limiting the quality factor to $\lesssim 10^3$?}

\cor{To answer this question, here we investigate high-impedance NbTiN resonators in a variety of dielectric configurations that are commonly used when fabricating double-quantum dots based on semiconductor nanowires}~\cite{fasth2005tunable,petersson2012circuit,hu2012hole,barker2019individually,froning2021ultrafast}.
We demonstrate that despite a reduction of the resonator quality factors due to the additional dielectrics, their quality is sufficient even in a dielectric configuration ideal for nanowire integration.

We describe the crucial parameters during sputtering of the material and investigate the dominant resonator loss mechanism.
A crucial criterion for resonators aiming on coupling to spin qubits is their magnetic-field resilience which can be achieved by employing disordered superconductors with a high critical magnetic field~\cite{samkharadze2016high,kroll2019magnetic,zollitsch2019tuning,borisov2020superconducting,yu2021magnetic}. Recently the community has started to operate spin qubits at elevated temperatures~\cite{petit2020universal,yang2020operation,camenzind2022hole}.
Using these arguments as a motivation, we characterize the resonator performance in large magnetic fields up to 5\,T and at elevated temperatures. 
\section{experimental setup}
We fabricated a total of 16 NbTiN coplanar waveguide resonators with an impedance of $Z=\sqrt{L/C}\sim 2$\,k$\Omega$ distributed on four different chips with differing dielectric configurations.
The sample preparation is described in section~\ref{sec:sampleprep}.
Each chip hosts a feedline with four notch-type, half-wave resonators as shown in Fig.~\ref{fig:hangers}a).
From left to right, the different wrapping of the feedline results in coupling quality factors between $Q_c\approx10^3$ and $Q_c\approx10^5$ \cor{(see Table~\ref{tab:my_label} in the appendix). This large spread of $Q_c$ enables} us to investigate the film properties accurately for a large range of internal quality factors $Q_i$. Resonance frequencies are in the range between $f_r\sim 4.2$\,GHz and $f_r\sim5.6$\,GHz.

Fig.~\ref{fig:hangers}b) shows the four different dielectric configurations in colored, dashed boxes.
For illustrative purposes, the corresponding dielectric configuration of a nanowire device is shown as well \cor{and we will explain the advantages and disadvantages of each configuration in the context of nanowire integration below.}
Case$~$I - NbTiN on Si - is the most ideal configuration for the resonator. 
But fabricating bottom-gate based devices~\cite{froning2021ultrafast,froning2018single} directly on top of intrinsic silicon comes with the problem of gate leakage, because of the small gate pitch.
Moreover, on the surface of the intrinsic silicon, a native silicon oxide forms under ambient conditions which might result in a poor electrostatic device stability.
Therefore, fabrication of nanowire devices on top of intrinsic silicon involves sandwiching the bottom gates with two oxide layers grown by atomic-layer deposition (ALD).
To maintain a pure dielectric environment of the resoantor, the oxide has to be wet-etched or the ALD-layers have to be deposited locally by a lift-off process~\cite{biercuk2003low}.
Wet-etching of the oxide might lead to unwanted surface-chemistry on the surface of the NbTiN~\cite{toomey2018influence}.
And, since ALD growth is a conformal processes, the lift-off process might result in irregular, rough edges around the desired structures that may protrude significantly out of plane with respect to the substrate (red crosses in schematic).
These edges in turn, may lead to step coverage issues on subsequent metal layers.

The local deposition of oxides for nanowire device integration is alleviated if the whole chip, including the resonator, can be covered with an ALD-grown oxide.
We investigate this in case$~$II - Al\textsubscript{2}O\textsubscript{3} on NbTiN on Si. 

For device integration, it is desirable to work with electrostatically silent oxides.
Therefore, nanowire devices are commonly fabricated on top of thermally grown silicon oxide.
We therefore investigate the performance of resonators on top of silicon oxide in case III - NbTiN on SiO\textsubscript{2} on Si.
In this case bottom-gate based nanowire devices only require one local oxide deposition step as indicated in the schematic.
Additionally, the remaining local oxide deposition is alleviated in case IV -  Al\textsubscript{2}O\textsubscript{3} on NbTiN on SiO\textsubscript{2} on Si.

The color codes as introduced in Fig.~\ref{fig:hangers}b) are used throughout the rest of the manuscript and denote the dielectric configuration and Tab.~\ref{tab:participationratios} shows the calculated participation ratios of each dielectric for the four investigated dielectric configurations.
\begin{table}
\centering
\resizebox{\linewidth}{!}{\begin{tabular}{l|cccc}
\hline\hline
&\textcolor{red}{NbTiN} & \textcolor{purple}{NbTiN+Al$_2$O$_3$} & \textcolor{blue}{SiO$_2$+NbTiN} & \textcolor{cyan}{SiO$_2$+NbTiN+Al$_2$O$_3$}\\ 
      \hline
vacuum & 0.0963     & 0.0933       & 0.1016 & 0.0947 \\
Al$_2$O$_3$ (14\,nm)  & -     & 0.0127           & -                       & 0.0215                      \\
SiO$_2$ (100\,nm)  & -     & -           & 0.2035                      & 0.1830                     \\
Si (500\,\textmu m)     & 0.9038    & 0.8940          & 0.6949                      & 0.7009\\
\hline\hline
\end{tabular}
}
\caption{\cor{\textbf{Participation ratios.} Fraction of
electric field energy stored in a each dielectric layer for the four investigated films. Values are obtained from performing a dc finite-element simulation using COMSOL.}}
\label{tab:participationratios}
\end{table}
\section{Sample preparation\label{sec:sampleprep}}
Since this work aims to investigate resonator losses due to the choice of the dielectric configuration, the intrinsic $Q$ of the resonators  must not be limited by the NbTiN film quality. Here we summarize the steps taken to optimize the fabrication of the used films.

1. As a substrate, we select two undoped Si wafers with a resistivity larger than 10\,k$\Omega$cm; one with only a layer of native SiO$_2$ and the other with $\sim$100\,nm of thermally grown SiO$_2$.

2. In order to minimize the impurity density at the metal-substrate interface, the wafer with only native oxide undergoes the following etching steps: (i) a Piranha etch to oxidize the top $\sim$10\,nm that may contain contamination, (ii) an HF bath to remove this oxide layer, and (iii) a second Piranha etch, followed by (iv) a second HF bath seconds before loading the wafer into the sputtering chamber. The second wafer hosts $\sim$100\,nm of thermally grown SiO\textsubscript{2}. In order to remove organic residues but keeping the oxide layer intact, we consecutively use ultrasonic cleaning of the wafer in an aqueous solution of tripotassium orthophosphat~\footnote{The used solution has the brand name deconex\textsuperscript{\textregistered} 12 BASIC 2\% solution.}, distilled water, acetone and isopropanol before loading the wafer into the sputtering chamber.

3. The vacuum quality in the sputtering chamber plays a vital role. We perform Ti pre-sputtering, resulting in a significant reduction of the chamber base pressure.

4. We pre-sputter the NbTi target to remove the top, potentially contaminated or oxidized layer~\cite{vissers2010low}.

5. The sputtering rate has to be maximized by choosing an ideal set of sputtering parameters. See Appendix~\ref{appendix:sputter} for details.
Because the impingement rates of oxygen and water decreases with increasing growth rates, higher sputtering rates result in a purer film and accordingly lower loss tangents of the resonators~\footnote{Discussion with Mihai Gabureac}.

6. We perform sputtering as close as possible to stoichiometry of NbTiN~\cite{ohya2013room}. See Appendix~\ref{appendix:sputter} for details.

7. The resonators are dry-etched using argon/chlorine, offering a higher selectivity against silicon etching compared to the more widely used fluorine based etching recipes~\cite{vissers2010low,sandberg2012etch}.
This makes it easier to prevent over-etching.
\cor{
We note that etch-induced losses~\cite{sandberg2012etch}  might limit internal Q factors to $\sim 10^5$, beyond the regime of our interest.}

8. After fabrication, each film is characterized in dc measurements by measuring the critical temperature $T_c^\mathrm{dc}$ and the sheet resistance $R^\mathrm{sq,dc}$ close to $T_c^\mathrm{dc}$ using etched reference structures.
\cor{This allows us to \cor{estimate} the sheet kinetic inductance~\cite{tinkham2004introduction,annunziata2010tunable,hong2013terahertz} which we use to design the resonator geometry.}
The resonance frequency is designed using analytical equations of coplanar waveguide resonators~\cite{gupta1990microstrip} and the coupling quality factor is estimated by simulating the structure using the electromagnetic simulation software Sonnet.
\section{Determining loss due to two-level fluctuators}
\begin{figure}[ht]
    \centering
    \includegraphics[width=\linewidth]{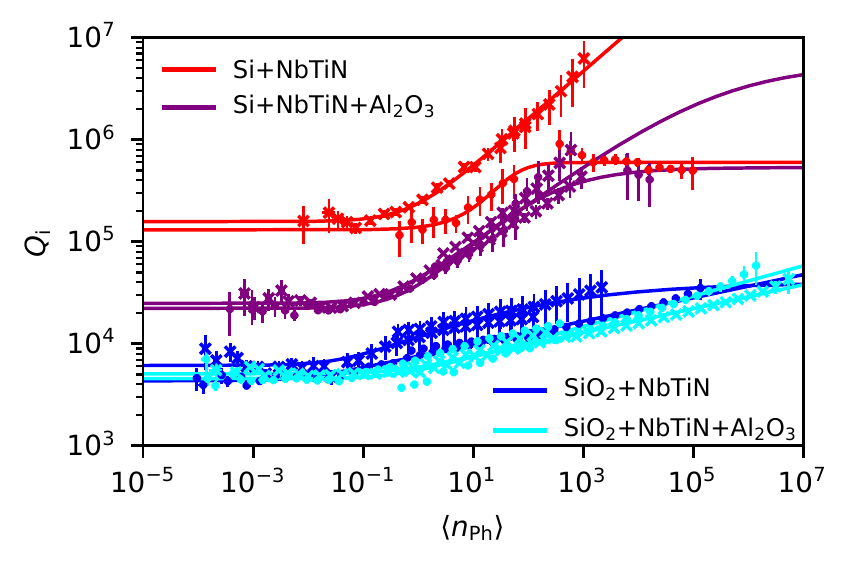}
    \caption{\textbf{Power dependence.} Internal quality factor $Q_i$ as a function of average photon number $\langle n_\mathrm{ph}\rangle$ in the resonator. The solid lines are fits to Eq.~\eqref{eq:QTLS}, assuming two level fluctuators as the dominating leakage mechanism at low photon numbers. \cor{For each dielectric configuration, as encoded by the color according to Fig.~\ref{fig:hangers}b), data for the two resonators with minimal $\left|Q_\mathrm{TLS}-Q_c\right|$ are plotted. } Different symbols correspond to different resonators.}
    \label{fig:power}
\end{figure}

\begin{table}[ht]
    \centering
    \resizebox{\linewidth}{!}{\begin{tabular}{l|cc|cc}
    \hline\hline
&\multicolumn{2}{c|}{film A: Si substrate}&\multicolumn{2}{c}{film B: SiO$_2$ substrate}\\
& \textcolor{red}{NbTiN} & \textcolor{purple}{NbTiN+Al$_2$O$_3$} & \textcolor{blue}{NbTiN} & \textcolor{cyan}{NbTiN+Al$_2$O$_3$}\\
\hline
$R^{\mathrm{sq,dc}}$ ($\Omega$)&\multicolumn{2}{c|}{$260\pm20$}&\multicolumn{2}{c}{$430\pm40$}\\
$T_c^\mathrm{dc}$ (K)&\multicolumn{2}{c|}{$5.8\pm0.1$}&\multicolumn{2}{c}{$6.6\pm0.3$}\\
$L_\mathrm{k}^\mathrm{sq,rf}$ (pH)&\multicolumn{2}{c|}{$79\pm14$}&\multicolumn{2}{c}{$61\pm9$}\\
$D$ (cm\textsuperscript{2}/s)&\multicolumn{2}{c|}{$0.27\pm0.08$}&\multicolumn{2}{c}{$0.32\pm0.33$}\\
$Q_\mathrm{TLS}$ ($10^3$)&$152\pm9$&$23.1\pm0.9$&$4.6\pm0.3$&$5.1\pm0.2$\\
$Q_\mathrm{other}$ ($10^6$)&$6.3\pm0.5$&$0.63\pm0.34$&-&-\\
$n_c$&$0.8\pm0.3$&$0.32\pm0.42$&$0.014\pm0.006$&$0.6\pm0.6$\\
$\beta$&$0.43\pm0.04$&$0.46\pm0.29$&$0.12\pm0.02$&$0.13\pm0.02$\\
\cor{
$Q_i^s$ $(10^3)$}& $80\pm 30$& $19 \pm 2$& $3.2 \pm 0.3$&
$2.6 \pm  0.3$\\
\cor{$\Delta_s$ (mT)}& $21\pm 7$&  $66\pm 8$&
$76\pm7$&
$76\pm 7$\\
\cor{$g_s$}& $1.98\pm 0.08$ & $1.94 \pm 0.06$ & $1.782 \pm 0.012$ & $1.72 \pm 0.04$\\
\hline\hline    \end{tabular}}
    \caption{\textbf{Resonator properties.}
    Extracted parameters for the two films with varying dielectric configurations.
    The sheet resistance $R^\mathrm{sq,dc}$ and critical temperature $T_c^\mathrm{dc}$ are obtained from a dc measurement.
    The rf sheet kinetic inductance $L_\mathrm{k}^\mathrm{sq,rf}$ is independently inferred from the 8 measured resonance frequencies of either film, where the error represents the root variance.
    $Q_\mathrm{TLS}$, $Q_\mathrm{other}$, $n_c$ and $\beta$ are fit parameters of Eq.~\eqref{eq:QTLS}.
    \cor{The paramagnetic impurities leading to the feature in Fig.~\ref{fig:Bfielddep}a) are characterized by the minimum internal Q-factor $Q_i^s$ and feature width $\Delta_s$. The values of $D$, $Q_\mathrm{TLS}$, $Q_\mathrm{other}$, $n_c$, $\beta$, $Q_i^s$, and $\Delta_s$ represent weighted averages over the data sets of 4 resonators of either dielectric configuration with weights proportional to the inverse of the error bar of the fit, resulting in a maximum weight, if $Q_c\sim Q_i$ for which the resonance is most pronounced. $g_s$ is the Landé g-factor extracted from Fig.~\ref{fig:Bfielddep}b).}
    The color code corresponds to Fig.~\ref{fig:hangers}b).\label{tab:maintable}}
\end{table}
\begin{figure}[tb]
    \centering
     \begin{overpic}[width=\linewidth]{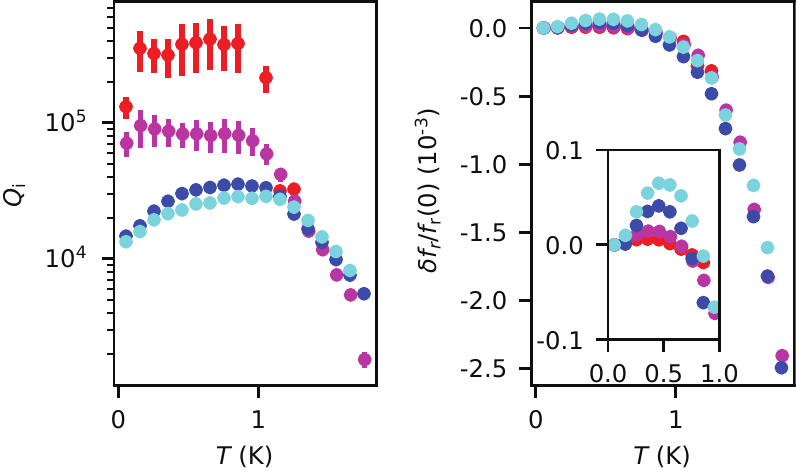}
\linethickness{1pt}
\put(0,56){a)}
\put(50,56){b)}
\end{overpic}
    \caption{\textbf{Temperature dependence of resonator properties.} a) Internal quality factor $Q_i$ as a function of temperature. b) Relative frequency shift $\delta f_r=(f_r(T)-f_r(0))$ as a function of temperature. The inset is a zoom in onto the peak that is observed at $\sim$ 0.5 K. In all sub-figures, the color encodes the dielectric configuration according to Fig.~\ref{fig:hangers}b) and legend in Fig.~\ref{fig:power}. The data was acquired at an average photon number of $\sim10^4$.}
    \label{fig:temperature} 
\end{figure}
To benchmark the performance of the resonators in the different dielectric configurations, we probe each notch-type half-wave resonator by measuring the transmission $S_{21}$ through the feedline at the base temperature of a dilution refrigerator $T_\mathrm{base}\approx 30$\,mK.
Fig.~\ref{fig:hangers}c) shows an exemplary resonance which is described by~\cite{khalil2012analysis,gao2008physics}
\begin{equation}
S_{21}=1 - \frac{Q_le^{i\Phi}}{Q_c\cos(\Phi) (1 + 2iQ_l(\omega/\omega_r-1))}\label{eq:S21hanger}.
\end{equation}
Here $Q_l=1/(Q_i^{-1}+Q_c^{-1})$ is the loaded quality factor and $\Phi$ describes a small resonance asymmetry due to interference with a standing-wave background~\cite{khalil2012analysis}.

We identify 4 resonances on every investigated chip and assign every measured resonance frequency $f_r$ to a physical resonator. 
Using the physical dimensions of the resonator for calculating its geometric inductance and capacitance~\cite{gupta1990microstrip}, we deduce its kinetic inductance from the measured resonance frequency. 
Thereby, we use the center conductor width which we measure for every resonator by means of scanning electron microscopy after having performed the experiments presented in this work.
The so obtained averaged square kinetic inductance $L_\mathrm{k}^\mathrm{sq,rf}$ is given in Table~\ref{tab:maintable}  where the error bar represents the root variance. Values for individual resonators are shown in Table~\ref{tab:my_label} in the appendix
We find that the values of $L_k^\mathrm{sq,dc}$ and $L_k^\mathrm{sq,rf}$ are \cor{consistent for} either film.

The differences in $R^\mathrm{sq,dc}$ and $T_\mathrm{c}^\mathrm{dc}$ between the two films such as the large variance of $L_\mathrm{k}^\mathrm{sq,rf}$ is attributed to the small film thickness giving rise to a large effect of film thickness inhomogeneities.

In order to quantify the loss due to two-level systems (TLS) residing in the differing dielectric structures, we measure a resonance trace for every resonator in every dielectric configuration and extract $Q_i$ as a function of power applied on the feedline by fitting Eq.~\eqref{eq:S21hanger} in a circular fit~\cite{probst2015efficient}. Fig.~\ref{fig:power} shows the fitted internal quality factor $Q_i$ for two resonators of each configuration. We convert the applied power $P_\mathrm{in}$ on the feedline to an average photon number in the resonator using~\cite{yu2021magnetic}
\begin{equation}
\langle n_\mathrm{ph}\rangle=\frac{Q_c}{\omega_r}\left(\frac{Q_i}{Q_i+Q_c}\right)^2\frac{P_\mathrm{in}}{\hbar\omega_r},
\end{equation}
where $Q_i$, $Q_c$ and $\omega_r$ are extracted from fitting the resonance curve. TLS residing in the oxides close to the resonator give rise to a power dependent dielectric loss which is usually modeled by~\cite{phillips1987two,wang2009improving,pappas2011two,goetz2016loss,brehm2017transmission,carter2019low,muller2019towards,scigliuzzo2020phononic}
\begin{equation}
\frac{1}{Q_i}=\frac{1}{Q_\mathrm{TLS}}\frac{\tanh\left(\frac{\hbar\omega_r}{2k_BT}\right)}{\left(1+\frac{\langle n_\mathrm{ph}\rangle}{n_c}\right)^\beta}+\frac{1}{Q_\mathrm{other}}\label{eq:QTLS}.
\end{equation}
In the low power limit, and at low temperatures, $Q_i$ is approximately given by $Q_\mathrm{TLS}$ due to TLS.
When increasing $\left< n_\mathrm{ph}\right>$ above a critical value $n_c$, $Q_i$ increases with a characteristic scaling $\beta$ until eventually saturating at $Q_\mathrm{other}$.
We fit Eq.~\eqref{eq:QTLS} to the data (solid lines in Fig.~\ref{fig:power}) and extract $Q_\mathrm{TLS}$, $Q_\mathrm{other}$, $n_c$ and $\beta$ as fit parameters. 
The weighted average of these fit parameters for each film are specified in Table~\ref{tab:my_label}.

We find that Eq.~\eqref{eq:QTLS} fits well to our data in all four dielectric configurations implying that in the limit of low photon numbers, all resonators are limited by their coupling to TLS.
However, the quantitative behavior 
for the different dielectric configurations differs by a lot.
Let us first consider the low-photon limit in Fig.~\ref{fig:power}. At low photon numbers, the internal quality factor is determined by the coupling to TLS, $Q_i (n=0)\sim Q_\mathrm{TLS}$.

In case I - NbTiN on Si, $Q_i$ saturates at the largest value as the number of photons in the resonator approaches zero. This implies a low abundance of TLS at the interface between the intrinsic silicon and the NbTiN. In case II - Al\textsubscript{2}O\textsubscript{3} on NbTiN on Si, $Q_i$ saturates at values approximately an order of magnitude lower which we attribute to the larger abundance of TLS stemming from the ALD-grown oxide on top of the metal \cor{and on top of the dielectric}.
For case III - NbTiN on SiO\textsubscript{2} - and for case IV - Al\textsubscript{2}O\textsubscript{3} on NbTiN on SiO\textsubscript{2}, the saturation of $Q_i$ in the low-photon limit happens another order of magnitude lower than for case II. We attribute this decrease to the larger participation ratio (compare Tab.~\ref{tab:participationratios} of the layers below the center conductor compared to the ones above it due to the larger dielectric constant of silicon as compared to the vacuum dielectric constant.
The larger importance of the oxides below the center conductor is confirmed by the negligible difference of $Q_\mathrm{TLS}$ in case III and case IV (with additional oxide on top of the resonator).

In all cases, once the average number of photons $\langle n_\mathrm{ph}\rangle$ exceeds a critical value $n_c$, $Q_i$ increases, because the TLS are increasingly saturated and no longer open a photon leakage path~\cite{phillips1987two,pappas2011two,sage2011study,goetz2016loss}. 

In the high power limit in case I and case II, all TLS saturate, and $Q_i$ asymptotically approaches $Q_\mathrm{other}$ which originates from a power independent source of loss.
The origin of $Q_\mathrm{other}$ potentially lies in the interaction with phonons or quasiparticles.
In case III and case IV, $Q_i$ does not saturate even at photon numbers on the order of $10^7$ underlining the importance of losses due to TLS in these cases.

Despite the TLS being the dominant source of loss for these resonators, we highlight that $Q_\mathrm{TLS}$ well exceeds $10^3$ even for the configuration where the resonator is sandwiched between SiO\textsubscript{2} and Al\textsubscript{2}O\textsubscript{3}.
This result is a central point of this manuscript as it allows for easier integration of semiconductor nanowires into a resonator architecture maintaining a good resonator quality.
Moreover, we stress that $Q_\mathrm{TLS}$ is larger by almost an order of magnitude when oxides are only grown on top of the metal and not below.
\section{Resonator stability at elevated temperatures and fields}
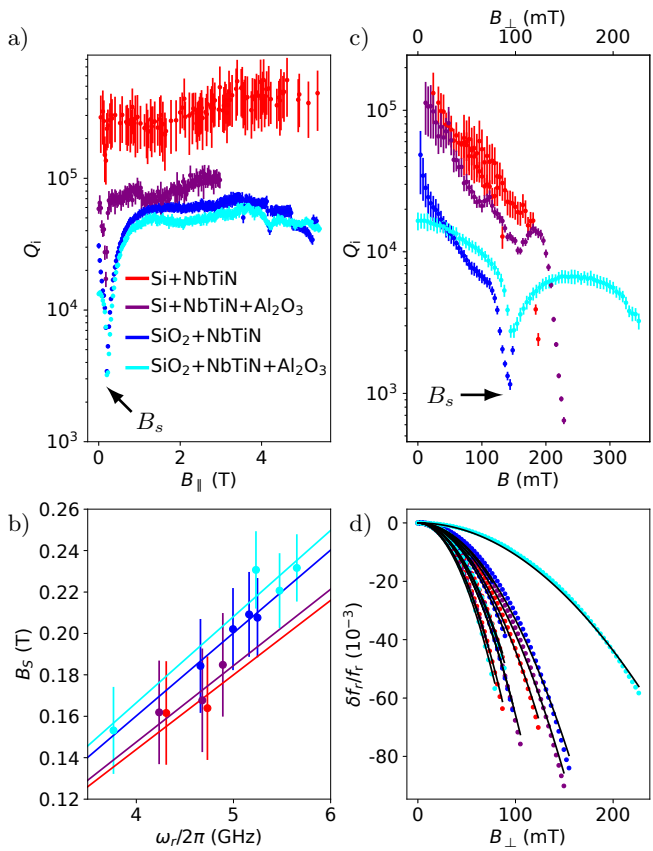
\begin{figure}[ht]
    \centering
    \begin{overpic}[width=\linewidth]{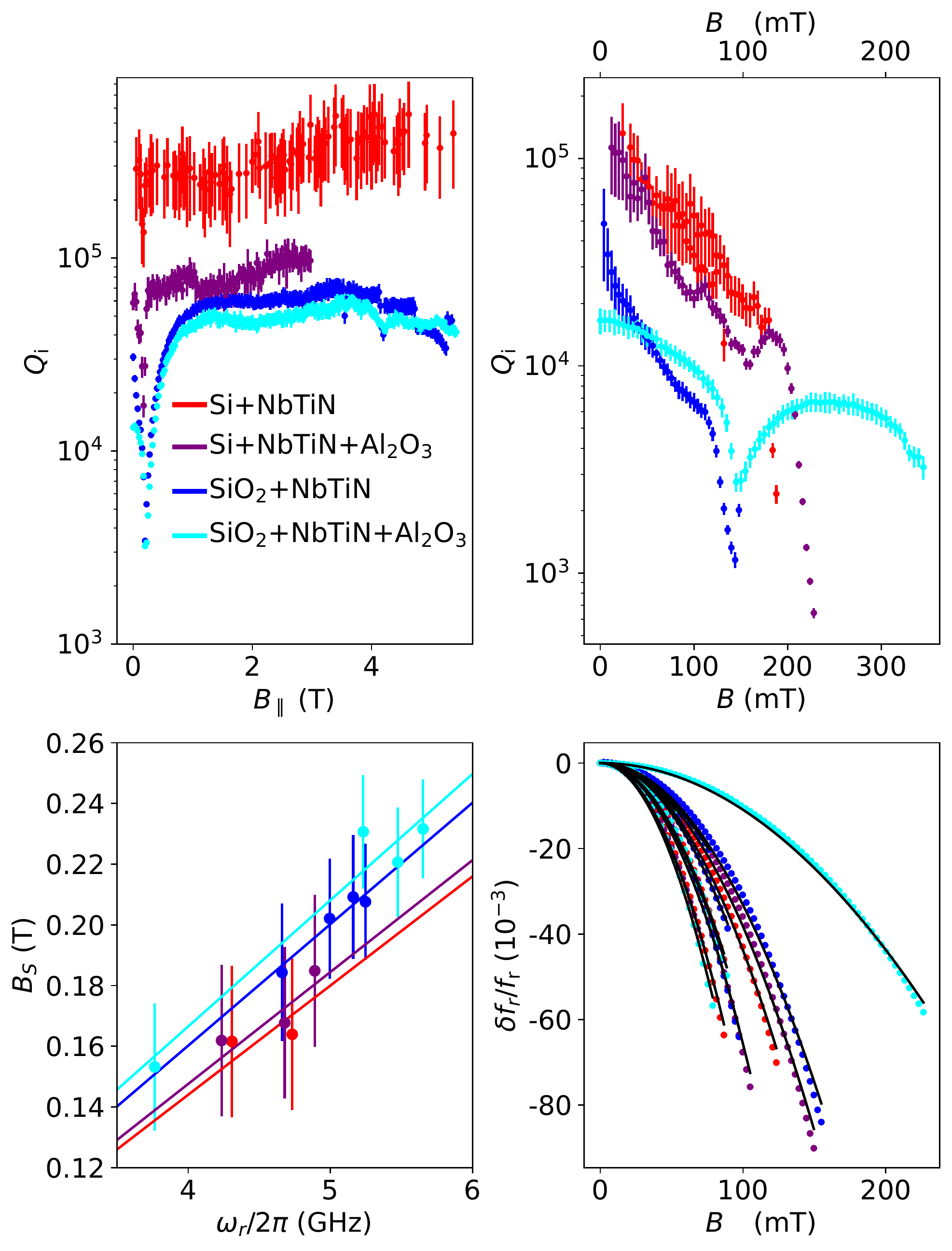}
\linethickness{1pt}
\put(0,95){a)}
\put(0,38){b)}
\put(40,95){c)}
\put(40,38){d)}
\put(14.5,52){\vector(-1,1){3}}
\put(15,50){$B_s$}
\put(53,54){\vector(1,0){5}}
\put(49,53){$B_s$}
\put(58,1){\tiny$\perp$}
\put(58,97){\tiny$\perp$}
\end{overpic}
    \caption{\cor{\textbf{Magnetic field dependence} a) Internal quality factor $Q_i$ as a function of in-plane field amplitude $B_\parallel$. A dip is observed at a field $B_s$ which is attributed to resonant paramagnetic impurities. b) $B_s$ extracted from data similar to the one shown in a) for different resonators as a function of resonance frequency $\omega_r$. The solid lines are fits to $B_s=\hbar \omega_r/g_s\mu_B$ from which we extract the Landé g-factor $g_s$ given in Table~\ref{tab:maintable}.}
    c) Internal quality factor $Q_i$ as a function of out-of-plane field. The field $B$ is applied with an angle of $49^\circ$ with respect to the substrate and the perpendicular component $B_\perp$ is indicated. Once again a dip is observed around $B=hf_r/2\mu_B$, being less pronounced for the resonator fabricated on intrinsic silicon. c) Relative frequency shift $\delta f_r=f_r(B_\perp)-f_r(0)$ as a function of out-of-plane field component $B_\perp$. The solid lines are fits to Eq.~\eqref{eq:Bperpdep}. In all sub-figures, the color encodes the dielectric configuration according to Fig.~\ref{fig:hangers}b).}
    \label{fig:Bfielddep}
\end{figure}
After having determined the quality of the resonators in each dielectric configuration, quantified by $Q_\mathrm{TLS}$, we aim on benchmarking the resonator stability at elevated temperatures and magnetic fields in regimes relevant for spin-qubit operation.

Fig.~\ref{fig:temperature}a) shows $Q_i$ as a function of temperature for all dielectric configurations where the color code corresponds to the one introduced in Fig.~\ref{fig:hangers}b).
For all curves, we measure an increase in $Q_i$ for increasing temperature peaking at $\sim0.8$\,K. We attribute this increase in the quality factor to an increasing saturation of the TLS with temperature.
When the temperature exceeds $\sim1$\,K, a decline in $Q_i$ is observed which is attributed to an increasing quasiparticle population because of the closing of the superconducting gap~\cite{coumou2012microwave,samkharadze2016high}. 

Simultaneously to measuring the quality factor, we also measure the shift in resonance frequency $\delta f_r$ and plot it in Fig.~\ref{fig:temperature}b).
We find that $\delta f_r$ peaks at a temperature $T_0\sim 0.5$\,K (see inset of figure), with the effect being most pronounced for the resonators fabricated on top of SiO\textsubscript{2} (blue points in Fig.~\ref{fig:temperature}).
Moreover, the positive frequency shift at increased temperature  exceeds the positive frequency shift in the case of TLS saturation due to a large photon population (see Fig.~\ref{fig:freq_shift_vsN} in the appendix). This effect can be understood by noting that the latter only saturates TLS in a narrow band around the resonance frequency $\omega_r$ while the elevated temperature saturates TLS in a much wider frequency range~\cite{pappas2011two}.
The temperature of maximum positive frequency shift corresponds to a frequency of $\omega_\mathrm{thermal}/2\pi=k_\mathrm{B}T_0/h\sim 10$\,GHz.
As $\omega_\mathrm{thermal} > \omega_r$, saturation of TLS in this frequency range explains the postitive resonance frequency shift due to the dispersive interaction between resonator and TLS. For larger temperature the resonance frequency starts to decrease due to the closing of the superconducting gap $\Delta$ resulting in a larger kinetic inductance, explaining the decrease of $\delta f_r$.

In order to benchmark the stability of the NbTiN resonators in an external magnetic field, we plot $Q_i$ as a function of in-plane field $B_\parallel$ in Fig.~\ref{fig:Bfielddep}a) and as a function of out-of-plane field $B_\perp$ in Fig.~\ref{fig:Bfielddep}b).
During the latter measurement, the magnetic field $B$ was applied with an angle of $49^\circ$ with respect to the sample plane as indicated by the second longitudinal axis in Fig.~\ref{fig:Bfielddep}b).
In both cases, we observe a dip in $Q_i$ at an absolute field strength $B_s$.
The dip is attributed to a resonant interaction with paramagnetic impurities in the substrate~\cite{samkharadze2016high}. Remarkably, the dip is much less pronounced for the resonator fabricated without any additional oxides, indicating that the paramagnetic impurities mainly reside within the oxides.
\cor{We quantitatively analyze the influence of the paramagnetic impurities by fitting a lorentzian to the dip of each resonator. Therefrom, we extract the minimum quality factor $Q_i^s$, the feature width $\Delta_s$ and the field $B_s$ around which the dip is observed (see Table~\ref{tab:my_label} in the appendix). Fig.~\ref{fig:Bfielddep}(b) shows $B_s$ versus the resonance frequency of the corresponding resonator $\omega_r$. This allows to extract the Landé g-factor $g_s$ by fitting a straight line given by $B_s=\hbar\omega_r/g_s\mu_B$, where $\mu_B$ is the Bohr magnetron. The resulting values of $g_s$ for the different dielectric configurations are shown in Table~\ref{tab:maintable}.}

Besides this dip-like feature, for the in-plane field, no noteworthy decline in $Q_i$ is observed up to the very largest applied field strengths of 5\,T, confirming a magnetic-field resilience for in-plane fields.
This is expected because the penetration depth $\lambda\sim260$\,nm~\cite{hong2013terahertz} is much larger than the thickness of the NbTiN film ($\sim10$\,nm).
Out-of-plane, $Q_i$ declines monotonously for increasing field-strengths which we attribute to an increasing quasiparticle density in the film.
However, $Q_i$ remains larger than $10^3$ up to $B_\perp\sim 100$\,mT for all dielectric configurations, once again confirming their suitability for spin qubit integration.

Finally, Fig.~\ref{fig:Bfielddep}c) shows the resonance frequency versus out-of-plane field.
The data is well fitted by
\begin{equation}
\frac{\delta{f_r}}{f_r(0)}=-\frac{\pi}{48}\frac{De^2}{\hbar k_BT_c}w^2 B_\perp^2 \label{eq:Bperpdep}
\end{equation}
which is deduced from BCS theory~\cite{samkharadze2016high,tinkham2004introduction}.
Here, the width $w$ of every resonator center conductor is measured by electron beam microscopy. The average width is $w=390\pm120$\,nm where the error bar is the root variance. $D$ denotes the diffusion constant which is a fit parameter.

The weighted average of the fitted diffusion constants are given in Table~\ref{tab:maintable} and the values for individual resonators are given in Table~\ref{tab:my_label}. 
We note that the diffusion constants of the two investigated films are similar and slightly lower than the one in Reference~\cite{samkharadze2016high}.

\section{Conclusion}
We have investigated superconducting, high-impedance resonators based on NbTiN in four different dielectric configurations. The largest internal quality factor in the low-photon limit is found for the resonator fabricated on intrinsic silicon.
Nevertheless, all other dielectric configurations result in internal quality factors $>10^3$, \cor{a value which has proven sufficient for performing key experiments in the context of spin qubits.}
Moreover, we benchmark the resonator performance at elevated temperatures and magnetic-field strengths.
Since the resonators are compatible with existing fabrication protocols, our results allow for straightforward integration of these types of resonators with various kinds of spin qubits defined in semiconductor nanowires. 

We acknowledge very fruitful discussions with Mihai Gabureac, Sergey Amitonov and Alessia Pally. Furthermore, we thank Dario Marty for the support in wafer etching in the facilities of the Paul Scherrer Institute. This research was supported by the Swiss Nanoscience Institute (SNI), the Swiss National Science Foundation through grant 192027 and through the NCCR Spin Qubit in Silicon (NCCR-Spin). We further acknowledge funding from the European Union’s Horizon 2020 research and innovation programme, specifically the FET-open project AndQC, agreement No 828948 and the FET-open project TOPSQUAD, agreement No 847471. We also acknowledge support through the Marie Skłodowska -Curie COFUND grant QUSTEC, grant agreement N° 847471 and the Basel Quantum Center through a Georg H. Endress fellowship.
All data
in this publication are available in numerical form at:
\url{https://doi.org/10.5281/zenodo.7602078}.
\bibliographystyle{apsrev4-1} 
 \bibliography{bib}\cleardoublepage
 \appendix
 \section{Investigation of sputtering parameters\label{appendix:sputter}}
\renewcommand\thefigure{A.\arabic{figure}}    \setcounter{figure}{0}
 \begin{figure}[!htb]
    \centering
    \includegraphics[width=\linewidth]{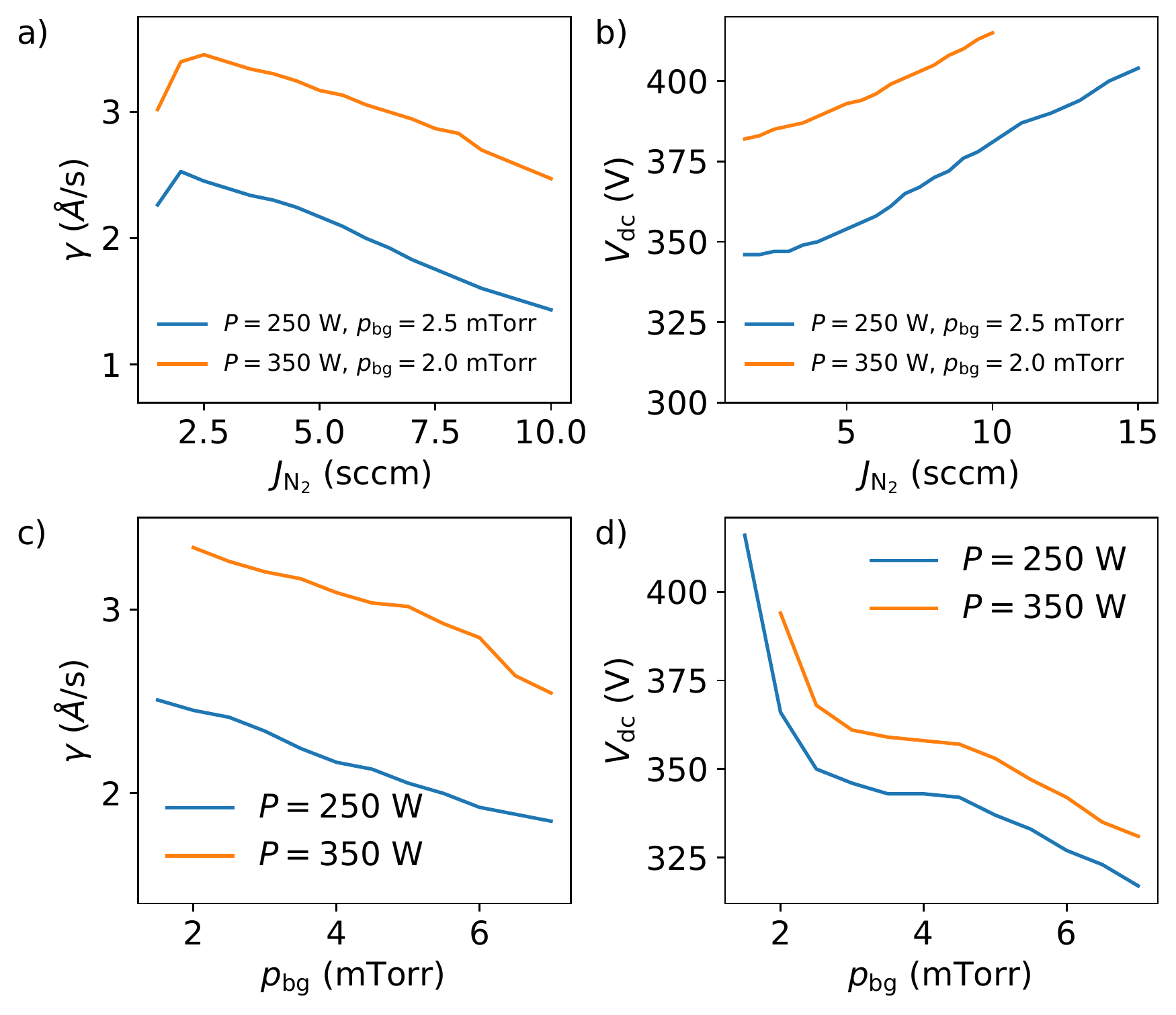}
    \caption{\textbf{Sputtering parameters.} a) growth rate $\gamma$ as function of nitrogen flow $J_{N_2}$ at different back-ground pressures $\rho_\mathrm{bg}$ and plasma power $P$.
    b) dc voltage $V_\mathrm{dc}$ between plasma source and target as function of $J_{N_2}$ for two valued of $\rho_\mathrm{bg}$ and $P$. A kink in the curve is obersed in both a) and b) at the same values of $J_{N_2}$. This kink corresponds to sputtering at a stoichiometric ratio. c) growth rate $\gamma$ as a function of background pressure $\rho_\mathrm{bg}$ showing a monotonous decay as lower $\rho_\mathrm{bg}$ correspond to larger mean-free paths and therefore to a smaller scattering of the sputtered material. d) dc voltage $V_\mathrm{dc}$ between the plasma source and the target as a function of background pressure $\rho_\mathrm{bg}$. When lowering $\rho_\mathrm{bg}<2$\,mTorr, $V_\mathrm{dc}$ increases drastically and the plasma becomes unstable. All traces were measured at a fixed argon flow rate $Q_\mathrm{Ar}=50$\,sccm.}
    \label{fig:sputtering}
\end{figure}
In this appendix, we give detailed background information about the choice of sputtering parameters that are used while fabricating the resonators as described in Section~\ref{sec:sampleprep} in the main text.

In order to minimize the impurity density of the sputtered NbTiN film, it is desirable to maximize the growth rate $\gamma$, because a shorter sputtering time results in less gathered contaminants in the film. While sputtering, the plasma power $P$, the background pressure $\rho_\mathrm{bg}$, the argon flow $J_\mathrm{Ar}$ and the nitrogen flow $J_{\mathrm{N}_2}$ can be controlled. In Fig.~\ref{fig:sputtering}, we investigate the dependence of the growth rate $\gamma$ and the voltage between the plasma source and the target $V_\mathrm{dc}$ on these parameters. The growth rate increases as a function of $P$. Therefore, the power should be chosen as high as possible while maintaining a stable plasma which is the case in our chamber for $P\lesssim250$\,W.
As a function of $J_{\mathrm{N}_2}$, a maximum in the growth rate is found, corresponding to the stoichiometric ratio~\cite{glowacka2014development} (see Fig.~\ref{fig:sputtering}a) and Fig.~\ref{fig:sputtering}b)).
The position of the optimum depends on $P$ and $\rho_\mathrm{bg}$. When increasing $\rho_\mathrm{bg}$, $\gamma$ decreases (see Fig.~\ref{fig:sputtering}c)). Therefore, the background pressure should be chosen as small as possible before the plasma becomes unstable. In our sputtering chamber, this is the case for $\rho_\mathrm{bg}\lesssim2$\,mTorr. We choose $P=250$\,W, $\rho_\mathrm{bg}=2$\,mTorr, $J_\mathrm{Ar}=50$\,sccm, $J_{\mathrm{N}_2}=3.5$\,sccm for the sputtering of both films.
\section{Frequency shift in power dependence\label{appendix:shift}}
\begin{figure}[b]
    \centering
    \includegraphics[width=\linewidth]{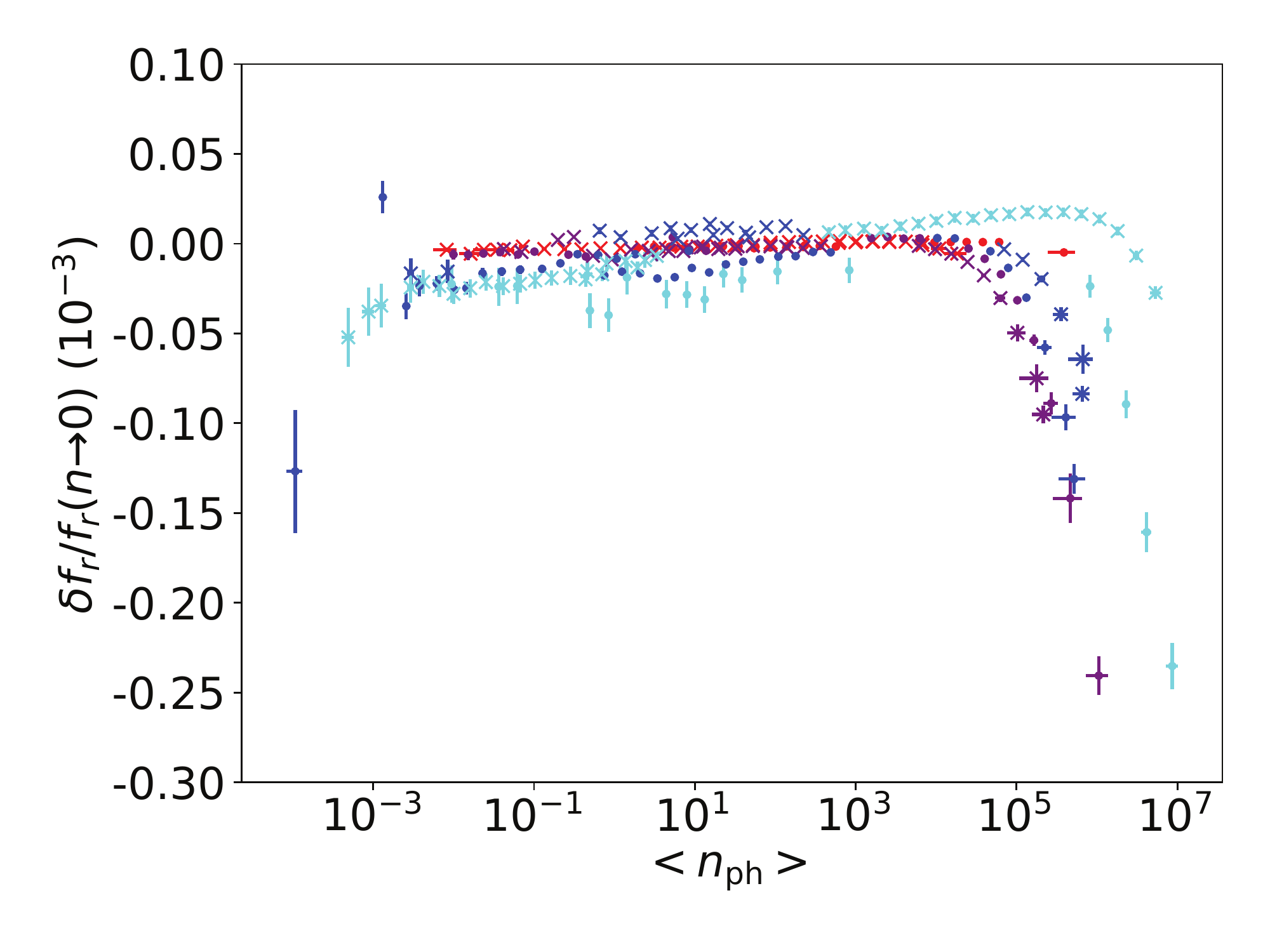}
    \caption{\textbf{Power dependence.} Relative frequency shift $\delta f_r=f_r-f_r(n\to0)$ as function of average number of photons in the resonator $\langle n_\mathrm{ph}\rangle$. We do not observe a noteworthy positive frequency shift in comparison with the Fig.~\ref{fig:temperature} in the main text. For $\langle n_\mathrm{ph}\rangle\gtrsim10^5$, an increasingly negative relative frequency shift is observed which we attribute to the finite non-linearity of the resonators.}
    \label{fig:freq_shift_vsN}
\end{figure}
In the inset of Fig.~\ref{fig:temperature}b) in the main text, we observe a substantial positive resonance frequency shift as a function of temperature with a peak at approximately 0.5\,K.
We attribute this shift to a saturation of TLS with a transition frequency of approximately 10\,GHz dispersively interacting with the resonator.
Fig.~\ref{fig:freq_shift_vsN} shows the resonance frequency shift as a function of number of photons in the resonator.
The observed positive shift is smaller by an order of magnitude compared to Fig.~\ref{fig:temperature}. 

For very large drive powers with $\langle n_\mathrm{ph}\rangle \gtrsim10^5$, a negative relative frequency shift is observed in Fig.~\ref{fig:freq_shift_vsN}. We attribute this negative shift to the onset of the bifurcation of the resonator due to a finite non-linearity.
\section*{Additional data}
All extracted data for each of the 16 individual resonators is shown in Table~\ref{tab:my_label}.
\begin{table*}[htbp]
    \centering
    \begin{tabular}{l|cccc}
    &\multicolumn{4}{c}{\textcolor{red}{Si + NbTiN}}\\
    &resonator \#1&resonator \#2&resonator \#3&resonator \#4\\
\hline    
$f_r$ (GHz)  &$ 4.308 $&$ 4.728 $&$4.732 $&$4.807 $\\
$L_\mathrm{k}^\mathrm{sq,rf}$ (pH)&$93.5$&$62$&$77$&$75$\\
$Q_c$ &$(3800\pm400)$&$(17000\pm3000)$&$(119000\pm9000)$&$(400\pm40)$\\
$D$ (cm\textsuperscript{2}/s) &$0.21$&$0.39$&$0.26$&$0.29$\\
$Q_\mathrm{TLS}(10^{5})$&-&$(1.15\pm0.24)$&$(1.6\pm0.08)$&$(1.6\pm1.2)$\\
$Q_\mathrm{other}(10^{5})$&-&$(6.3\pm0.5)$&-&$(1.87\pm0.08)$\\
$n_c$&-&-&$(1.0\pm0.4)$&-\\
$\beta$&-&$(1.7\pm1.5)$&$(0.49\pm0.06)$&-\\
$Q_i^S(10^{4})$&$(4.62\pm0.08)$&-&$(9.01\pm0.04)$&-\\
$\Delta_S$&-&-&$(0.021\pm0.007)$&-\\
\hline
&\multicolumn{4}{c}{\textcolor{purple}{Si+NbTiN+Al$_2$O$_3$}}\\
&resonator \#5&resonator \#6&resonator \#7&resonator \#8\\
$f_r$ (GHz)  &$4.235$&$ 4.680$&$4.891$&$4.944$\\
$L_\mathrm{k}^\mathrm{sq,rf}$ (pH)&$104$&$68$&$80$&$74$\\
$Q_c$ &$(5700\pm700)$&$(100000\pm20000)$&$(12100\pm800)$&$(180\pm70)$\\
$D$ (cm\textsuperscript{2}/s) &$0.17$&$0.36$&$0.21$&$0.26$\\
$Q_\mathrm{TLS}(10^{4})$&$ (2.30\pm0.12)$&$(1.87\pm0.21)$&$(2.51\pm0.16)$&$(5.2\pm1.0)$\\
$Q_\mathrm{other}(10^{5})$&$(5.3\pm0.6)$&$(3.4\pm2.0)$&-&$(0.73\pm0.05)$\\
$n_c$&$(1.1\pm0.4)$&$(0.10\pm0.09)$&$(0.42\pm0.23)$&-\\
$\beta$&$(0.68\pm0.09)$&$(0.31\pm0.08)$&$(0.42\pm0.07)$&-\\
$Q_i^S(10^{4})$&-&$(1.866\pm0.026)$&$(3.40\pm0.04)$&-\\
$\Delta_S$&-&$(0.065\pm0.008)$&$(0.07\pm0.018)$&-\\
\hline
&\multicolumn{4}{c}{\textcolor{blue}{Si +SiO\textsubscript{2} + NbTiN}}\\
&resonator \#9&resonator \#10&resonator \#11&resonator \#12\\
$f_r$ (GHz)  &$4.660$&$4.99$&$5.161$&$5.247$\\
$L_\mathrm{k}^\mathrm{sq,rf}$ (pH)&$74$&$53$&$60$&$54$\\
$Q_c$ &$(10000\pm900)$&-&$(19000\pm2000)$&$(2200\pm300)$\\
$D$ (cm\textsuperscript{2}/s) &$0.25$&$0.44$&$0.31$&$0.41$\\
$Q_\mathrm{TLS}(10^{3})$&$(4.33\pm0.33)$&-&$(4.34\pm0.29)$&$(6.2\pm0.6)$\\
$Q_\mathrm{other}$&-&-&-&-\\
$n_c$&$(0.013\pm0.010)$&-&$(0.011\pm0.007)$&$(0.026\pm0.024)$\\
$\beta$&$(0.116\pm0.019)$&-&$(0.107\pm0.014)$&$(0.16\pm0.04)$\\
$Q_i^S(10^{3})$&$(3.76\pm0.08)$&$(2.33\pm0.06)$&$(3.73\pm0.07)$&$(4.78\pm0.08)$\\
$\Delta_S$&$(0.071\pm0.015)$&$(0.085\pm0.012)$&$(0.084\pm0.013)$&$(0.065\pm0.011)$\\
\hline
&\multicolumn{4}{c}{\textcolor{cyan}{Si +SiO\textsubscript{2} + NbTiN+Al\textsubscript{2}O\textsubscript{3}}}\\
&resonator \#13&resonator \#14&resonator \#15&resonator \#16\\
$f_r$ (GHz)  &$3.765$&$5.231$&$5.474$&$5.652$\\
$L_\mathrm{k}^\mathrm{sq,rf}$ (pH)&$60$&$73$&$61$&$51$\\
$Q_c$ &$(4500\pm330)$&$(130000\pm20000)$&$(14000\pm2000)$&$(2200\pm200)$\\
$D$ (cm\textsuperscript{2}/s) &$1.22$&$0.15$&$0.22$&$0.33$\\
$Q_\mathrm{TLS}(10^{3})$&$ (4.13\pm0.32)$&$(5.0\pm0.5)$&$(4.93\pm0.223)$&$(10.9\pm0.7)$\\
$Q_\mathrm{other}(10^{4})$&-&-&-&$(5.1\pm0.5)$\\
$n_c$&$(0.41\pm0.32)$&$(1.1\pm1.0)$&$(0.36\pm0.14)$&$(3.0\pm1.9)$\\
$\beta$&$(0.152\pm0.032)$&$(0.119\pm0.029)$&$(0.118\pm0.013)$&$(0.81\pm0.26)$\\
$Q_i^S (10^{3})$&$(3.81\pm0.05)$&$(1.07\pm0.11)$&$(2.93\pm0.04)$&$(4.11\pm0.10)$\\
$\Delta_S$&$(0.062\pm0.009)$&$(0.15\pm0.05)$&$(0.081\pm0.011)$&$(0.119\pm0.022)$\\
\hline
    \end{tabular}\caption{\cor{\textbf{Parameters for all 16 resonators} The shown parameters are the resonance frequency $f_r$, the sheet kinetic inductance $L_k^\mathrm{sq,rf}$, the coupling Q factor $Q_c$, the diffusion constant $D$, the low-power internal Q factor $Q_\mathrm{TLS}$. $n_c$ is the critical photon number and $\beta$ the scaling parameter from Eq.~\eqref{eq:QTLS}. $Q_i^S$ is the internal Q factor on resonance with the paramagnetic impurities (compare Fig.~\ref{fig:Bfielddep}(a)) and $\Delta_S$ is the width of this resonance. Values are missing where the data could not be extracted.}
    }
    \label{tab:my_label}
\end{table*}
\end{document}